\documentclass[sigconf, screen, 9pt]{acmart}
\AtBeginDocument{%
  \providecommand\BibTeX{{%
    \normalfont B\kern-0.5em{\scshape i\kern-0.25em b}\kern-0.8em\TeX}}}
\copyrightyear{2025}
\acmYear{2025}
\setcopyright{rightsretained}
\acmConference[ASPDAC '25]{30th Asia and South Pacific Design Automation
Conference}{January 20--23, 2025}{Tokyo, Japan}
\acmBooktitle{30th Asia and South Pacific Design Automation Conference
(ASPDAC '25), January 20--23, 2025, Tokyo, Japan}
\acmDOI{10.1145/3658617.3697778}
\acmISBN{979-8-4007-0635-6/25/01}

\begin{document}
\title{Compilation for Dynamically Field-Programmable Qubit Arrays with Efficient and Provably Near-Optimal Scheduling}

\author{Daniel Bochen Tan}
\email{bochentan@ucla.edu} \orcid{0000-0002-9711-2441}
\affiliation{
  \institution{University of California, Los Angeles}
  \city{Los Angeles}
  \state{CA 90095}
  \country{USA}}
\affiliation{
  \department{Department of Physics}
  \institution{Harvard University}
  \city{Cambridge}
  \state{MA 02138}
  \country{USA}}
\author{Wan-Hsuan Lin}
\email{wanhsuanlin@ucla.edu} \orcid{0000-0002-7486-2143}
\affiliation{
  \institution{University of California, Los Angeles}
  \city{Los Angeles}
  \state{CA 90095}
  \country{USA}}
\author{Jason Cong}
\email{cong@cs.ucla.edu} \orcid{0000-0003-2887-6963}
\affiliation{
  \institution{University of California, Los Angeles}
  \city{Los Angeles}
  \state{CA 90095}
  \country{USA}}
\renewcommand{\shortauthors}{Tan, Lin, and Cong}

\begin{abstract}
Dynamically field-programmable qubit arrays based on neutral atoms feature high fidelity and highly parallel gates for quantum computing.
However, it is challenging for compilers to fully leverage the novel flexibility offered by such hardware while respecting its various constraints.
In this study, we break down the compilation for this architecture into three tasks: scheduling, placement, and routing.
We formulate these three problems and present efficient solutions to them.
Notably, our scheduling based on graph edge-coloring is provably near-optimal in terms of the number of two-qubit gate stages (at most one more than the optimum).
As a result, our compiler, Enola, reduces this number of stages by 3.7x and improves the fidelity by 5.9x compared to OLSQ-DPQA, the current state of the art.
Additionally, Enola is highly scalable, e.g., within 30 minutes, it can compile circuits with 10,000 qubits, a scale sufficient for the current era of quantum computing.
Enola is open source at \href{https://github.com/UCLA-VAST/Enola}{https://github.com/UCLA-VAST/Enola}

\end{abstract}

\maketitle
\newtheorem{thm}{Theorem}
\renewcommand{\thethm}{\arabic{thm}}

\section{Introduction} \label{sec:intro}

In recent years, quantum computing based on neutral atoms has advanced quickly in scale, quality, and adoption.
Large experiments exceed 1,000 qubits~\cite{Pause:24-1000qubits}, at the forefront of quantum computing.
One-qubit gates with 99.97\% fidelity and two-qubit gates with 99.5\% fidelity have been demonstrated~\cite{nature22-lukim-bluvstein-atom-array, evered2023highfidelity} to be competitive among the platforms.
As a result, in addition to intensifying academic efforts, multiple startup companies~\cite{quera, infleqtion, pasqal, planqc, atomcomputing} have been established to pursue this route of quantum computing.

A particular advantage of neutral atoms is the ability to move the qubits.
Via these movements, the coupling among qubits is field-programmable dynamically in different stages of the quantum circuit execution.
This allows for a lot more flexibility to apply two-qubit entangling gates that are essential to quantum computing.
Thus, researchers were able to run some of the most advanced quantum circuits requiring non-local connectivity on the \textit{dynamically field-programmable qubit arrays} (DPQA) architecture~\cite{nature22-lukim-bluvstein-atom-array, bluvstein2023logical}.

\begin{figure}[t]
    \centering
    \includegraphics[scale=1]{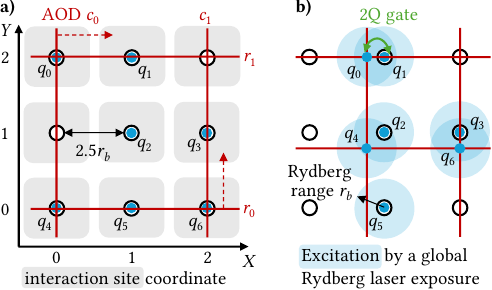}
    \caption{
        Dynamically field-programmable qubit arrays (DPQA).
        a) Qubits (blue dots) can transfer between SLM traps (circles) and AOD traps (intersections of red lines).
        AOD rows and columns can move while preserving their relative order.
        b) A global Rydberg laser excites all qubits.
        A two-qubit gate is applied if two qubits are within the Rydberg range.
    }
    \label{fig:dpqa}
\end{figure}

In DPQA, qubits are captured in two kinds of traps.
A spatial light modulator (SLM) generates an array of \textit{static} traps, as indicated by the 3-by-3 circles in Fig.~\ref{fig:dpqa}.
Seven of these traps are occupied by qubits.
A 2D acousto-optic deflector (AOD) generates \textit{mobile} traps that can travel in the plane.
The AOD traps are intersections of a set of rows and columns.
In our example, there are two rows ($r_0$ and $r_1$) and two columns ($c_0$ and $c_1$).
When we align the AOD traps with SLM traps and ramp up the AOD intensity, qubits are \textit{transferred} from the SLM to the AOD.
In Fig.~\ref{fig:dpqa}a, three qubits ($q_0$, $q_4$, and $q_6$) get transferred to the AOD.
Then, the AOD row $r_0$ shifts upward while the AOD column $c_0$ shifts to the right, so that the qubits in the AOD move along with them.
This movement yields the new configuration shown in Fig.~\ref{fig:dpqa}b.
At this point, if we reverse the movement and wind down the AOD, the three qubits would be transferred back to the SLM.
A major constraint of the movements is that the order of AOD columns cannot change, e.g., $c_0$ cannot move past $c_1$ to the right side, nor can the order of rows.
An order violation may cause the qubits in the AOD to collide and be lost.

\begin{figure*}[t]
    \centering
    \includegraphics[scale=1]{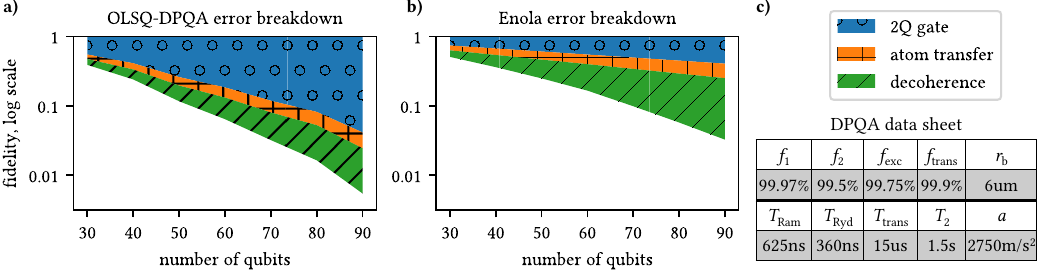}
    \caption{
        Error breakdown of the OLSQ-DPQA and Enola results.
        The benchmarks are 3-regular MaxCut QAOA circuits used in Ref.~\cite{tan2023compiling}.
        For the 90-qubit circuits, Enola reduces two-qubit gate stages by 3.7x and improves the overall fidelity by 5.9x.
    }
    \label{fig:breakdown}
\end{figure*}

A global Rydberg laser, which \textit{excites all qubits} to potentially entangle with each other, induces the multi-qubit interaction in DPQA.
The range of this interaction is named the \textit{Rydberg range}, $r_b$, illustrated by the half-transparent blue spheres in Fig.~\ref{fig:dpqa}b.
If two and only two qubits are within $r_b$ of each other, a controlled Z-rotation, e.g., a CZ gate, is applied.
In our example, three gates are applied: $(q_0,q_1)$, $(q_2,q_4)$, and $(q_3,q_6)$.
We call these parallel gates induced by the Rydberg laser a Rydberg \textit{stage} in the circuit execution.
Between these stages, qubits can be rearranged to different \textit{interaction sites} to interact with different qubits.
In Fig.~\ref{fig:dpqa}a, these sites are represented by the gray regions.
They center at integer points in the coordinate system and are separated sufficiently by 2.5$r_b$ so that multi-qubit interactions can only happen between qubits at the same site.
Note that, even if a qubit is alone during a Rydberg stage so that it does not go through a gate, such as $q_5$ in Fig.~\ref{fig:dpqa}b, it still gets excited by the Rydberg laser and accumulates error at the same rate as if it were involved in a gate.
Therefore, we should minimize the number of stages to reduce these side effect errors.

Assuming the quantum circuit to execute has been decomposed into a proper gate set supported by DPQA, the compilation then involves a few tasks: \textit{scheduling} gates to stages, \textit{placement} of the qubits, and \textit{routing} qubits between stages.
For fixed architectures, these three tasks are often formulated together as the circuit placement~\cite{circuit-placement, fan_QLSML_2022}, qubit mapping~\cite{asplos19-li-ding-xie-sabre-mapping, dac19-wille-burgholzer-zulehner-mapping-minimal-swaph, iccad21-tan-cong-qubit-mapping-absorption, huang2022reinforcement, li2023single_qubit_gate_matter_for_qls, park2022fsqm, asplos21-zhang-hayes-qiu-jin-chen-zhang-time-optimal-mapping, micro22-tannu-maxsat-qubit-mapping}, or quantum layout synthesis (QLS) problem~\cite{tc20-tan-cong-optimality-layout-queko, iccad20-tan-cong-optimal-layout-synthesis, dac23-olsq2, wu_robust_2022,  ShaikvdP2023_qls_as_classical_planning}.
OLSQ-DPQA~\cite{iccad22-olsq-raa, tan2023compiling} is the first work to investigate QLS for DPQA with an effort to find the optimal QLS solutions.
However, it can only handle circuits with up to 90 qubits \textit{in a day} because it depends on solving satisfiability modulo theories (SMT) problems, which is NP-complete.
The long compilation time hinders the adoption of this approach.
Subsequent works Q-Pilot~\cite{wang2023qpilot} and Atomique~\cite{wang2024atomique} aim to develop scalable heuristic solutions for better scalability. 
However, as indicated by the authors of these heuristic works themselves, the result quality is notably worse than OLSQ-DPQA.

In this work, we present a compiler Enola (\underline{e}fficient and \underline{n}ear-\underline{o}ptimal \underline{l}ayout synthesizer for \underline{a}tom arrays) that is scalable and significantly improves result fidelity of previous works.
In Enola, a scheduler first assigns quantum gates to stages.
Then, a placer decides the location of qubits in each stage.
Finally, a router derives the detail movement instructions between stages.
On the 90-qubit 3-regular MaxCut QAOA (quantum approximate optimization algorithm) benchmarks~\cite{tan2023compiling}, Enola reduces the number of stage by 3.7x and improves the overall fidelity by 5.9x compared to OLSQ-DPQA.
In terms of scalability, we demonstrate compiling circuits with up to 10,000 qubits in 30 minutes, which is sufficient for all DPQA hardware available now or in the near future.

The fidelity improvements of Enola are mainly due to the reduction of Rydberg stages, as indicated by the suppression of the two-qubit gate portion in Fig.~\ref{fig:breakdown}b compared to Fig.~\ref{fig:breakdown}a.
Specifically, we can model two-qubit gates as edges in a graph so that assigning gates to stages becomes coloring the edges in the graph.
Suppose the optimal number of Rydberg stages is $S_\text{opt}$.
Leveraging the efficient and provably near-optimal Misra-Gries edge-coloring algorithm~\cite{misra_constructive_1992}, Enola manages to schedule the gates to $S_\text{opt}$ or $S_\text{opt}+1$ stages.
To our knowledge, we are the first to relate the DPQA scheduling problem to graph edge-coloring and leverage this near-optimal algorithm.
In comparison, the formulation of OLSQ-DPQA hinders the exploration of multiple rearrangement steps between two stages, resulting in much more stages than the optimum.

In Enola, the placement problem is solved by simulated annealing to reduce the qubit traveling distance, and the routing solution is generated by independent sets to avoid AOD order violations while facilitating parallelism.
To our knowledge, we are also the first to formulate the DPQA routing problem to solving independent sets.

The paper is organized as follows.
Sec.~\ref{sec:motivation} motivates the paper by analyzing the OLSQ-DPQA results with a detailed fidelity model.
The following three sections (\ref{sec:scheduling}, \ref{sec:placement}, and \ref{sec:routing}) provide our solutions of scheduling, placement, and routing.
Next, Sec.~\ref{sec:evaluation} presents the evaluations.
Then, Sec.~\ref{sec:related} introduces related works.
Finally, in Sec.~\ref{sec:conclusion}, we conclude the paper and suggest directions for future improvements that could address some limitations of this work.

\section{Motivation: Fidelity Analysis}\label{sec:motivation}


We model three error sources: imperfect gates, atom transfers, and qubit decoherence.
The qubit movements contribute to qubit idling time, which is accounted for in the decoherence term.
The parameters follow leading experiments~\cite{nature22-lukim-bluvstein-atom-array, bluvstein2023logical} and are summarized in Fig.~\ref{fig:breakdown}c.
Single-qubit gates have fidelity $f_1=99.97\%$ and duration $T_\text{Ram}=625\text{ns}$.
These gates can be individually addressed to corresponding qubits, so there are no side effect errors on other qubits.
In this work, we make the same assumption as Ref.~\cite{tan2023compiling} that the single-qubit gates are first removed so that \textit{the compiler only handles the two-qubit gates}.
Then, the single-qubit gates are inserted back to the compiled results.
Two-qubit gates have fidelity $f_2=99.5\%$ and duration $T_\text{Ryd}=360\text{ns}$.
The other qubits also excited by the Rydberg laser, e.g., $q_5$ in Fig.~\ref{fig:dpqa}a, each has the fidelity $f_\text{exc}=99.75\%$.
Atom transfers have fidelity $f_\text{trans}=99.9\%$ and duration $T_\text{trans}=15\text{us}$.
Note that multiple transfers can be simultaneous, e.g., the three transfers in Fig.~\ref{fig:dpqa}a will take 15us.
The coherence time of qubits is $T_2=1.5\text{s}$.
The decoherence effect of a qubit $q$ is modelled by a multiplicative factor $1-T_q/T_2$ where $T_q$ is its idling time, i.e., the total duration of the procedure carried out on DPQA minus any time spent on gates or transfers.
The majority of $T_q$ is spent on AOD movements.
The move distance, $d$, and time, $t$, follow the relation $d/t^2=a=2750\text{m/s\textsuperscript{2}}$~\cite{nature22-lukim-bluvstein-atom-array}, e.g., if $d=110\text{um}$, then $t=200\text{us}$.

The overall fidelity is computed by
\begin{equation}
    f=(f_{1})^{g_1} \cdot \underbrace{(f_{2})^{g_2} \cdot (f_\text{exc})^{|Q| S - 2g_2}}_\text{two-qubit gate} \cdot \underbrace{(f_\text{trans})^{N_\text{trans}}}_\text{atom transfer} \cdot \underbrace{\Pi_{q\in Q}\ (1-T_q/T_2)}_\text{decoherence}, \label{eq:fidelity}
\end{equation}
where $g_1$ and $g_2$ are the number of single-qubit and two-qubit gates, respectively, $Q$ is the set of qubits, $S$ is the number of stages, $|Q|S-2g_2$ calculates the qubits affected by the Rydberg laser but does not perform a gate, and $N_\text{trans}$ is the total number of atom transfers.
Since we only focus on two-qubit gates, the term $(f_1)^{g_1}$ is a constant for every compiler and we ignore it from now on.
As an example, we calculate the fidelity for the process in Fig.~\ref{fig:dpqa}.
There are 3 two-qubit gates so $(f_{2})^{g_2}=0.9950^3=0.9851$.
Only $q_5$ is excited but does not perform a gate so $(f_\text{exc})^{|Q|S - 2g_2}=f_\text{exc}^{7\times 1 - 2\times 3}=0.9975$.
Thus, the total \textit{two-qubit gate term} is $0.9851\times0.9975=0.9826$.
Since there are 3 atom transfers in Fig.~\ref{fig:dpqa}a, the \textit{atom transfer term} is $(f_\text{trans})^{N_\text{trans}}=0.9990^3=0.9970$.
The longest movement belongs to $q_4$: it travels a $\sqrt 2$ site separation, i.e., $\sqrt 2 \times 2.5r_b=21.21\text{um}$.
Thus, the AOD movement from Fig.~\ref{fig:dpqa}a to Fig.~\ref{fig:dpqa}b takes $t=(21.21\text{um}/2750\text{m/s\textsuperscript{2}})^{0.5}=87.82\text{us}$.
This is the $T_q$ for the moving qubits $q_0$, $q_4$, and $q_5$.
The other 4 qubits are additionally idling during the atom transfer, so their $T_q=87.82\text{us}+T_\text{trans}=102.82\text{us}$.
Therefore, the \textit{decoherence term} is $[1-87.82/(1.5\times 10^6)]^3\times [1-102.82/(1.5\times 10^6)]^4=0.9996$.
Finally, the overall fidelity is $f=0.9826\times 0.9970 \times 0.9996=97.92\%$.

In Ref.~\cite{tan2023compiling}, OLSQ-DPQA compiles a set of QAOA circuits designed for the MaxCut problem on 3-regular graphs~\cite{arxiv1411-farhi-goldstone-gutmann-qaoa} with the number of qubits ranging from 30 to 90.
The specific circuit is the problem unitary in QAOA, $U_C$, consisting of 3 commutable two-qubit gates on each qubit.
We evaluate their compiled results with our fidelity model and present the breakdown in Fig.~\ref{fig:breakdown}a.
Note that, to draw the figure, we take the logarithm of the fidelity terms so that they are additive.
At 90 qubits, the two-qubit gate fidelity term is 0.0414, the atom transfer term is 0.592, and the decoherence term is 0.223.
Thus, the dominating error source are the two-qubit gates, confirming claims Ref.~\cite{tan2023compiling}.
However, there is a gap between their number of stages, on average 14.6 for 90 qubits, and the theoretical lower bound, 3, because each qubit is only involved in 3 two-qubit gates.
Our compiler, Enola, produces only 4 stages, improving the two-qubit fidelity term to 0.406.
This effect is evident in the great decrease of the two-qubit gate portion in Fig.~\ref{fig:breakdown}b compared to Fig.~\ref{fig:breakdown}a.

\section{Scheduling: Edge Coloring}\label{sec:scheduling}

\begin{figure}[t]
    \centering
    \includegraphics[scale=1]{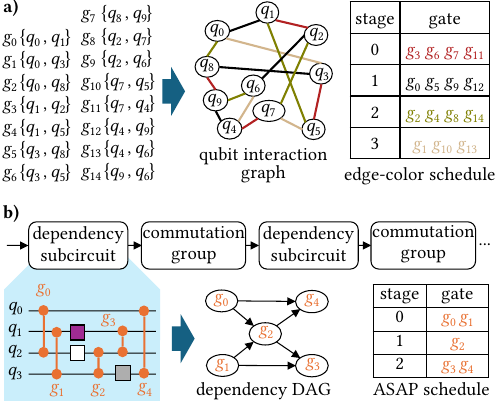}
    \caption{
        Scheduling in Enola.
        a) Scheduling a commutation group of two-qubit gates with edge coloring.
        b) Generic circuits can be divided to dependency subcircuits and commutation groups.
        Dependency subcircuits are scheduled ASAP.
    }
    \label{fig:scheduling}
\end{figure}

Since the fidelity term of two-qubit gate is more critical than atom transfer and decoherence, and previous methods are far from reaching the optimal number of stages, \textit{our metric in scheduling is the number of stages}.
We first consider the scheduling of a special class of circuit, \textit{commutation groups}, consisting of commutable two-qubit gates that can be executed in any order.
A commutation group can be represented by a \textit{qubit interaction graph} $G=(V,E)$ where the vertices are qubits and the edges are the two-qubit gates (Fig.~\ref{fig:scheduling}a).
The scheduling involves finding a function $\psi:E\to \mathbb{N}$ such that a qubit can only be involved in one gate at a Rydberg stage, i.e., for $e,e'\in E$ and $e\neq e'$, if $\psi(e)=\psi(e')$, then $e\cap e'=\emptyset$.

We now review some graph theory results.
For a graph $G=(V,E)$, an \textit{edge coloring} is a function $\phi:E\to \mathbb{Z}$ that evaluates different values for two different edges incident on a vertex, i.e., for $e,e'\in E$ and $e\neq e'$, if $e\cap e' \neq \emptyset$, then $\phi(e)\neq \phi(e')$.
Misra-Gries~\cite{misra_constructive_1992} provide an algorithm with runtime $O(|V|\cdot|E|)$ that gives an edge coloring $\Phi:E\to \{0,1,...,\Delta(G)\}$ where $\Delta(G)$ is the maximum vertex degree of $G$.
Thus, $\Phi$ colors the edges with at most $\Delta(G)+1$ colors.
The minimum number of colors to color the edges is called the \textit{chromatic index} of the graph, $\chi'(G)$.
Since $\chi'(G)\ge \Delta(G)$, $\Phi$ colors the edges with at most $\chi'(G)+1$ colors.
Our compiler leverages these results to schedule commutable two-qubit gates.
Its advantage is due to the following theorem.

\begin{thm}
    For a group of commutable two-qubit gates on $n$ qubits, suppose the optimal number of Rydberg stages to schedule these gates on DPQA is $S_\mathrm{opt}$, there is an algorithm with time complexity $O(n^3)$ that assigns these gates to at most $S_\mathrm{opt}+1$ Rydberg stages. 
\end{thm}

\textit{Proof.}
Note that the above definition of a schedule function, $\psi$, is contrapositive to the definition of an edge coloring function, $\phi$.
Thus, the scheduling function is just an edge coloring.
Therefore, the optimal number of Rydberg stages $S_\text{opt}=\chi'(G)$, which means the function $\Phi$ derived by the Misra-Gries algorithm maps the two-qubit gates to at most $S_\text{opt}+1$ Rydberg stages.
Since $|E|$ is $O(n^2)$ where $n$ is the number of qubits, and the Misra-Gries algorithm is $O(|V|\cdot|E|)$, the time complexity of our scheduling is $O(n^3)$.
\qed

A more generic quantum circuit is specified by a sequence of gates.
If two gates act on the same qubit, their relative order dictates a \textit{dependency}.
In Fig.~\ref{fig:scheduling}b, we exhibit an example of how one derives the dependency DAG (directed acyclic graph) for the two-qubit gates in a generic circuit.
In this case, the scheduling problem is straightforward: the optimal number of stages is the critical path in the DAG and ASAP (as soon as possible) scheduling can achieve optimality.
Although there is a way to augment the DAG to represent partially commutable circuits~\cite{iten-dag-commutation}, supporting this in general requires mixing logic synthesis and layout synthesis.
Therefore, we make an assumption similar to Ref.~\cite{yunong-commutation-group} that the whole quantum circuit is sliced into subcircuits that either respect all derived dependencies, as `dependency subcircuits' shown in Fig.~\ref{fig:scheduling}b, or are commutation groups.
The scheduling for the slices can be performed separately and the results can be stitched together afterwards.
This sliced structure is prevalent in quantum computing.
An example is the graph state preparation with various applications~\cite{book06-graph-state}, which has a layer of Hadamard gates in the beginning and then commuting CZ gates.
Another example is MaxCut QAOA that has alternating driver unitaries $U_B$ with dependency and problem unitaries $U_C$ that are commutation groups of ZZ gates.

In summary, for the dependency subcircuits, Enola uses ASAP scheduling which is optimal; for the commutation groups, Enola uses Misra-Gries algorithm, which is near-optimal.
Thus, the scheduling in Enola is near-optimal in the number of stages.

\section{Placement: Simulated Annealing}\label{sec:placement}

\begin{figure}[t]
    \centering
    \includegraphics[scale=1]{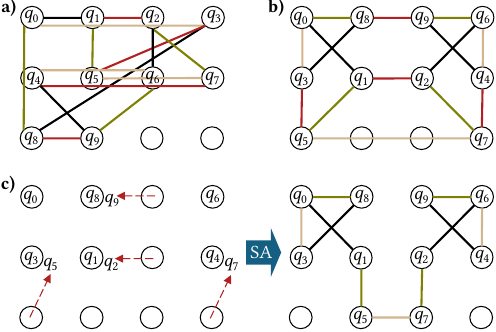}
    \caption{
        Placement in Enola.
        a) Trivial placement from left to right, from top to bottom.
        b) Placement with gate distance optimized by simulated annealing.
        c) Dynamic placement: after a Rydberg stage (red) is executed (left), run simulated annealing on moved qubits for a new placement (right).
    }
    \label{fig:placement}
\end{figure}

In placement, we map qubits to interaction sites.
The two-qubit gates at each Rydberg stage are like 2-pin nets in conventional circuit placement.
If a net has a long wire-length, it takes more time to move the qubits, resulting in more decoherence, the second largest error source.
Thus, we minimize the total wire-length in placement in order to reduce qubit movement time.
Note that this is heuristic because some qubits movements can be simultaneous.
As an example, qubits can be placed trivially from left to right and from top to bottom as in Fig.~\ref{fig:placement}a.
Then, the total distance of gates in the commutation group in Fig.~\ref{fig:scheduling}a accumulates to $25.67\times 2.5 r_b$.
In comparison, an optimized placement displayed in Fig.~\ref{fig:placement}b achieves a total wire-length of $19.48 \times 2.5 r_b$.
To minimize the qubit traveling distances, our cost function is defined as
\begin{equation}
    \sum\nolimits_{g(q,q')\in G} \ w_g\cdot \text{dist}(\ m(q),m(q')\ ),
\label{eq:placement-initial}
\end{equation}
where $w_g$ is the weight for gate $g$, $m$ is the circuit placement function from qubits to interaction sites, and `dist' is the Euclidean distance.
This is a common form of cost function in placement, and we apply a simulated annealing algorithm, Fast-SA~\cite{chen20006fastsa}, to optimize it.

To enhance exploration efficiency, we confine qubits to a specific region, thus reducing the search space.
Assuming the number of qubits to place is $n$, and the interaction sites have column indices $\{0,1,...,x_\text{max}\}$ and row indices $\{0,1,...,y_\text{max}\}$, we define the chip region for exploration as $x\in[0,\max(\lfloor\sqrt{n}\rfloor + 4, x_\text{max})]$ and 
$y\in[0,\max(\lfloor\sqrt{n}\rfloor + 4, y_\text{max})]$.
In Fig.~\ref{fig:placement}, $x_\text{max}=3$ and $y_\text{max}=2$.
Fast-SA has a three-stage annealing schedule to facilitate state space exploration.
At the first stage, the temperature is high.
In other words, we have higher probability to accept an inferior solution.
This stage mimics a random search to explore a large solution space.
Then, the second stage performs the pseudo-greedy local search with low temperature.
The last stage is a hill-climbing search where the temperature increases again to escape from local minima. 
The state in the annealing process is a placement which we initialize randomly.
Then, state transitions can be made by either reassigning a qubit to an empty site or exchanging the locations of two qubits.
The annealing process will terminate if the temperature is lower than a threshold or the number of iterations exceed a predefined limit, so the placement algorithm has a constant runtime.

The configuration after the first Rydberg stage (red) is on the left of Fig.~\ref{fig:placement}c.
The arrows indicate AOD movements from \ref{fig:placement}b to this configuration.
At this point, we can always reverse the movements to return to Fig.~\ref{fig:placement}b, and then find out the movements for the next stage (black).
In this case, the placement is static for all the Rydberg stages, so we set all the gate weights to 1 in the cost function.

However, one can also consider dynamically changing the placement for the next stage.
On the right of Fig.~\ref{fig:placement}c, we display a new placement where the gate between $q_5$ and $q_7$ is shorter compared to Fig.~\ref{fig:placement}b.
If the placement is dynamic, gates earlier in the schedule should contribute more to the cost function.
Thus, we set $w_g=\max(0.1, 1-0.1s_g)$, where $s_g$ is the number of stages preceding the stage that the gate $g$ belongs to, e.g., the gates in stages 0 to 3 will have the weights of 1, 0.9, 0.8, and 0.7, respectively.
During the simulated annealing for intermediate placement, we choose to move only the set of qubits necessitating relocation to vacant sites, while the remaining qubits stay where they are.
In our example, the new placement is from qubits $q_2$, $q_5$, $q_7$, and $q_9$ to the 6 empty sites.
The other qubits inherit the current placement.
We observe that dynamic placement outperforms static placement, so Enola uses dynamic placement by default.
In a commutation group, there are at most $O(n)$ stages, so our placement runtime is $O(n)$, given that each placement has a constant time limit.

\section{Routing: Independent Set}\label{sec:routing}

\begin{figure}[t]
    \centering
    \includegraphics[scale=1]{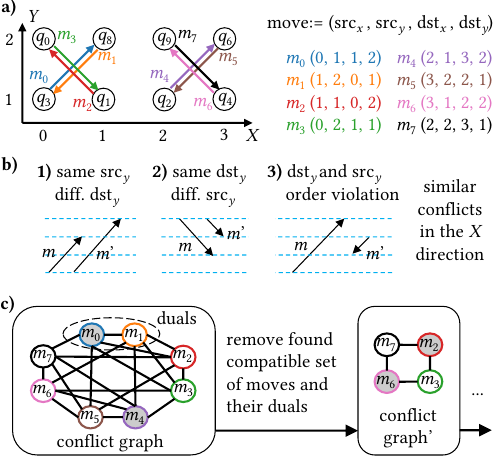}
    \caption{
        Routing in Enola.
        a) Definition of a move as a 4-tuple.
        b) Conflicts between two moves.
        c) Compatible moves are independent sets (IS) in the conflict graph (filled vertices).
        After finding an IS, delete the moves and their duals from the graph.
        The process continues until no moves are left.
    }
    \label{fig:routing}
\end{figure}

In routing, we would like to parallelize the AOD movements to reduce total execution time of quantum circuits.
Some movements cannot be performed simultaneously because of the fundamental constraints of AOD: \textit{the order of its columns cannot change, nor can the order of rows.}
We consider the movements of the second stage consisting of gates $(q_0, q_1)$, $(q_3,q_8)$, $(q_2,q_6)$, and $(q_4, q_9)$, in Fig.~\ref{fig:routing}a.
We define a \textit{move} to be a 4-tuple: $x$ and $y$ of the source, and $x$ and $y$ of the destination.
Since each gate has a choice of which one of its two qubits to move, there are two tuples corresponding to each gate.
For example, $m_0=(0,1,1,2)$ and $m_1=(1,2,0,1)$ are both for gate $(q_0, q_1)$.
We call them to be each other's \textit{dual}.
The AOD constraints are enforced by forbidding conflicts illustrated in Fig.~\ref{fig:routing}b.
If the sources of two moves $m$ and $m'$ have the same $y$, i.e., $\text{src}_y(m)=\text{src}_y(m')$, the two qubits are picked up by the same AOD row.
Then, $\text{dst}_y(m)=\text{dst}_y(m')$ because that AOD row can only terminate at one vertical location post-movement.
Similarly, if $\text{dst}_y(m)=\text{dst}_y(m')$, then $\text{src}_y(m)=\text{src}_y(m')$.
If the qubits are picked up by different rows, their relative order must be maintained, e.g., if $\text{src}_y(m)>\text{src}_y(m')$, then $\text{dst}_y(m)>\text{dst}_y(m')$.
In the $X$ direction, there are similar three types of conflicts.

These conflicts are pairwise, which means they can be encoded as edges in a graph where the vertices are the moves.
We present this \textit{conflict graph} in Fig.~\ref{fig:routing}c.
A set of compatible moves constitutes an independent set (IS) of vertices.
One can utilize a maximum independent set (MIS) solver for compatible moves, but MIS is NP-hard.\footnote{One can also imagine formulating the routing problem as a vertex coloring on the conflict graph, where each color represents a compatible set.
This formulation can find the optimal number of compatible sets whereas our approach above is greedy.
However, this method involves solving NP-hard coloring problems. Furthermore, since only one of the dual moves needs to be executed, this vertex coloring problem extends beyond the conventional definition.
Thus, we do not pursue this option in this work.}
In practice, we find \textit{maximal} independent sets are sufficient, which can be derived by a simple algorithm: 1) putting all vertices in a list, 2) adding the first vertex to the IS, 3) removing all its neighbors from the list, and continuing 2-3).
In the first box in Fig.~\ref{fig:routing}c, assuming the list of vertices is sorted by indices, $m_0$ is added to the IS first, so its neighbors $m_1$, $m_2$, $m_3$, $m_5$, and $m_7$ are removed from the list.
Next, $m_4$ is added to the IS and invalidates all the rest of vertices.
So, the maximal IS is $\{m_0, m_4\}$ corresponding to gates $(q_3,q_8)$ and $(q_2,q_6)$.
Next, $m_0$ and $m_4$, along with their duals $m_1$ and $m_5$ are deleted from the conflict graph, resulting in the second box in Fig.~\ref{fig:routing}c.
In the updated graph, we find the second maximal IS, $\{m_2,m_6\}$.
By now, all moves are deleted, and the routing terminates. 

The runtime of maximal IS is $O(|V|+|E|)$ where $|V|$ is the number of moves, which is less or equal than the number of qubits, $n$.
To construct the conflict graph, we need to check conflicts for all pairs of vertices, which requires $O(n^2)$ time.
The longest move in each compatible set determines the AOD movement time for this set.
Thus, in our compiler, the list of moves is sorted by their distance.
This sorting takes $O(n\log n)$ time.
Then, the maximal IS takes $O(n^2)$ time.
In summary, finding a compatible set takes $O(n^2)$ time.
In the worst case, each compatible set includes only one gate.
Then, we run $O(n)$ times the procedure above until all gates in one Rydberg stage are handled, resulting in $O(n^3)$ time.
In total, there can be $O(n)$ Rydberg stages for a commutable group, so the total routing time is $O(n^4)$.
We refer to this routing approach as \textit{sortIS}.

To improve the scalability of sortIS, we can introduce a fixed length window when scanning the possible moves.
Instead of constructing the whole conflict graph, we only construct a graph on the first $K$ vertices in the list where $K$ is the constant window size.
These vertices are the $K$ longest moves.
Thus, both checking the conflicts between vertices and solving the maximal IS only take $O(K^2)$ time.
Thus, the windowed routing takes $O(n^2 \log n + n^2K^2)$ time.
We refer to this routing approach as \textit{windowIS}.

For each compatible set of moves, the qubits need to be picked up by the AOD and dropped off to their destination interaction sites.
Turning on the AOD rows and columns and ramping up the intensity for atom transfers also take time.
To minimize this time, we need to consider the product structure of AOD, which is a research topic on its own~\cite{tan_depth-optimal_2024}.
In fact, it is proven that solving DPQA routing problems optimally is NP-hard from the complexity of optimizing the AOD pick-up time, as seen in Sec.~8.5 of Ref.~\cite{phd}.
In practice, we do not observe this subtask to be critical to the overall fidelity in the evaluations to follow.
Thus, in Enola, we adopt a simple approach implemented in a component named CodeGen in OLSQ-DPQA where the qubits are picked up row by row.
The columns may shift horizontally before picking up the next row.
The CodeGen just involves scanning over all the qubits to pick up, so the runtime is less than finding the compatible sets.

\begin{figure*}[t]
    \centering
    \includegraphics{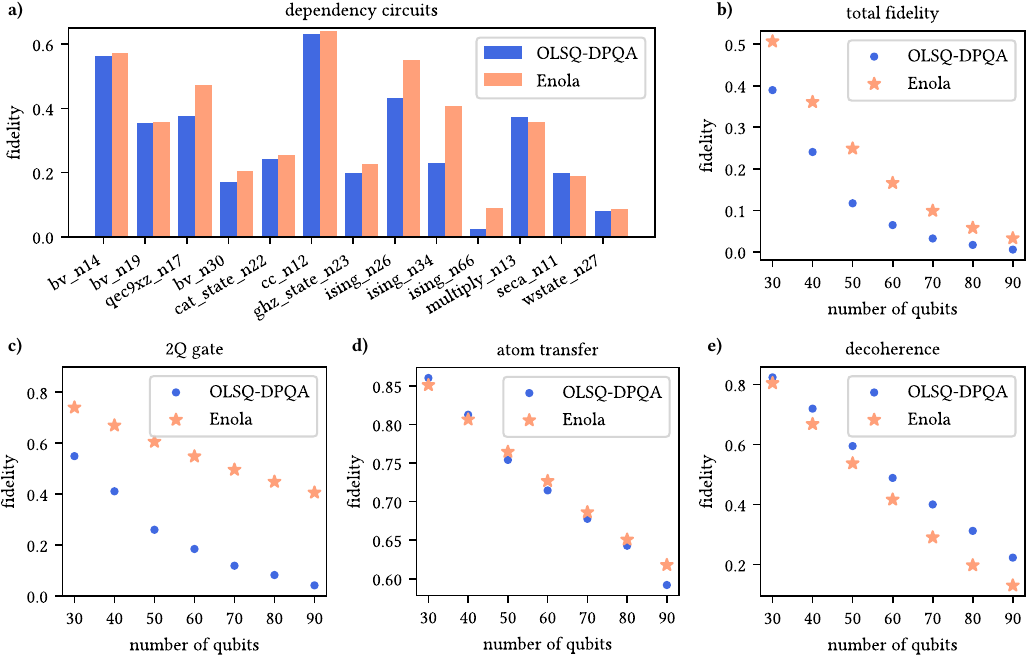}
    \caption{
        Result fidelity of Enola and OLSQ-DPQA on a) dependency circuits and b-e) 3-regular MaxCut QAOA circuits.
    }
    \label{fig:evaluation}
\end{figure*}

\section{Evaluation}\label{sec:evaluation}
We implemented our proposed algorithm in Python.
We employed KaMIS (v2.1)~\cite{DBLP:journals/jea/Hespe0S19} for solving the maximum independent set problems. 
All experiments were conducted on an AMD EPYC 7V13 64-Core Processor at 2450 MHz and 128 GB of RAM.
Each fidelity data point in the figures on QAOA is an average of results corresponding to 10 randomly generated graphs of the same size.

\subsection{Impact of Different Settings in Enola}
\begin{figure}[b]
    \centering
    \includegraphics[scale=1]{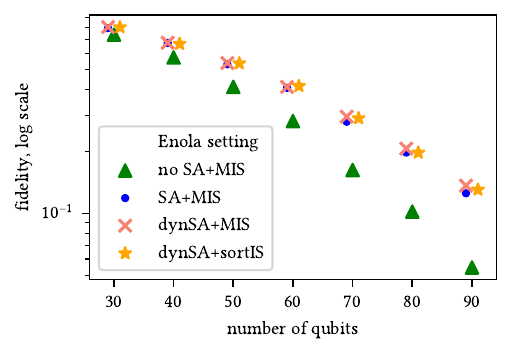}
    \caption{
        Decoherence fidelity term of different settings in Enola on 3-regular MaxCut QAOA circuits.
        `no SA' means trivial placement.
        `SA' means static placement.
        `dynSA' means dynamic placement.
        `MIS' means maximum independent using a solver.
        For these benchmarks, windowIS is the same with sortIS since the window size (1,000) is larger than the number of vertices in graph where we search for an IS.
    }
    \label{fig:settings}
\end{figure}

Fig.~\ref{fig:settings} provides the comparison of different settings in Enola on the MaxCut QAOA benchmarks.
Since the scheduling is the same for all settings, the two-qubit gate fidelity term is the same.
Additionally, in every setting, we use 4 atom transfers for each gate: picking up a qubit and dropping it off to the qubit it interacts with at this Rydberg stage, and the pick-up and drop-off on the way back.
This means the atom transfer fidelity term is also the same for all settings.
Thus, the comparison is on the decoherence term.
A major improvement comes from optimizing placement, as evident by the gap between trivial placement (green triangles) and the other series.
Dynamic placement (dynSA+MIS, pink cross) is slightly better than static placement (SA+MIS, blue dot).
In routing, sortIS is slightly worse than MIS, as in the comparison of dynSA+sortIS (yellow star) and dynSA+MIS (pink cross).
Thus, sortIS proves to be a viable replacement for MIS which is NP-hard.
The windowIS method is theoretically worse than sortIS because of the limited window size.
We set this size to be 1,000, larger than the scale of benchmarks in Fig.~\ref{fig:settings}.
In the evaluations with larger benchmarks up to 10,000 qubits, we observe a similar number of compatible move sets, and a similar average movement distance compared to sortIS, which means windowIS is a good heuristic to speed up the compilation.

\begin{figure}[t]
    \centering
    \includegraphics{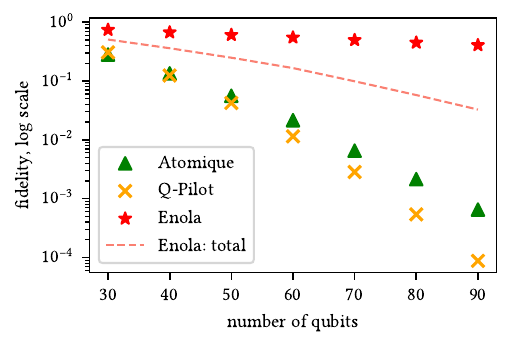}
    \caption{
        Comparison of two-qubit gate fidelity term (scattered markers) on 3-regular MaxCut QAOA circuits.
        The total fidelity of Enola is also drawn for reference (dashes).
    }
    \label{fig:comparison-heuristic}
\end{figure}

\subsection{Quality Comparison with Previous Works}
In Fig.~\ref{fig:evaluation}, we compare Enola with OLSQ-DPQA~\cite{tan2023compiling} using the same set of benchmarks in that study.
For dependency circuits\footnote{Out of the 45 benchmarks in Table 1 of Ref.~\cite{tan2023compiling}, we display the 13 highest-fidelity circuits in Fig.~\ref{fig:evaluation}a due to page limit. The observations we make do not change on the rest of the benchmarks.}, OLSQ-DPQA tries to execute as many gates as possible in the current front layer of the DAG, which often results in the same number of Rydberg stages as our ASAP scheduling.
In some cases, like the three `ising' benchmarks, OLSQ-DPQA suffers from elongating the critical path because its formulation cannot explore more than one rearrangement steps between Rydberg stages, which results in a notably worse fidelity compared to Enola.
In some other cases like `multiply\_n13' and `seca\_n11', it happens so that one rearrangement step is sufficient, so the two methods produce the same number of stages.
Under this scenario, OLSQ-DPQA can potentially outperform Enola because the routing in Enola is heuristic after all and may not find the optimal compatible sets of moves.

On the QAOA benchmarks, Enola clearly outperforms OLSQ-DPQA because it is able to leverage the near-optimal scheduling.
We depict the comparison of overall fidelity in Fig.~\ref{fig:evaluation}b and the three terms in Fig.~\ref{fig:evaluation}c-e.
In the two-qubit gate term, there is a significant gap between the two approaches.
At 90 qubits, OLSQ-DPQA uses 14.6 stages on average whereas Enola only employs 4, a 3.7x reduction.
In the atom transfer term, two approaches are similar, but Enola starts gaining advantage on larger benchmarks.
It should be noted that in OLSQ-DPQA, atom transfers are not penalized in the SMT formulation.
Examining its results with human eyes, there appears to be unnecessary transfers and movements.
In Enola, we only perform atom transfers and movements when necessary.
In the decoherence term, Enola is worse than OLSQ-DPQA.
This is inevitable because we choose to prioritize the number of Rydberg stages, necessitating more AOD movements.
Overall, Enola improves the fidelity by 5.9x compared to OLSQ-DPQA at 90 qubits.

The fidelity gain of Enola is even larger when compared to heuristic methods.
Q-Pilot~\cite{wang2023qpilot} is a DPQA router that utilizes AOD only for ancilla qubits to mediate two-qubit gates between SLM qubits.
It does not include a nontrivial placement solution.
Atomique~\cite{wang2024atomique} focuses on the placement and routes the qubits with SWAP gates.
In Fig.~\ref{fig:comparison-heuristic}, we compare the two-qubit gate fidelity terms of all the approaches.
Q-Pilot and Atomique result in much more Rydberg stages than Enola because the generation and recycling of the ancillas and the SWAPs require additional stages.
At 90 qubits, Enola reduces the number of stages by 8.7x compared to Atomique and 10.5x compared to Q-Pilot.
As a result, the two-qubit fidelity term of Enola (red star) is 779x higher than Atomique (green triangle) and 5806x higher than Q-Pilot (yellow cross).
The total fidelity of Enola (dashes), including atom transfers and decoherence, is still higher than the two-qubit fidelity term of the two heuristics.
Thus, the total fidelity of the heuristics is certainly worse than Enola.

\begin{figure}[t]
    \centering
    \includegraphics{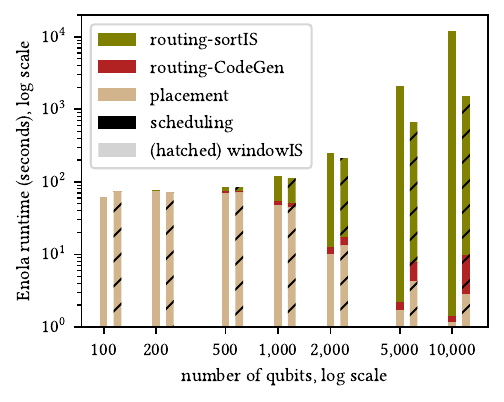}
    \caption{
        Enola runtime scaling on 3-regular MaxCut QAOA circuits.
        Two nearby bars correspond to the same number of qubits.
        The hatched bars are windowIS data (with a window size of 1,000) and the other bars are sortIS data. 
    }
    \label{fig:runtime}
\end{figure}

\subsection{Runtime Scaling of Enola}
Since the steps in Enola all have polynomial runtime, it is much more scalable than OLSQ-DPQA.
For 3-regular MaxCut QAOA, OLSQ-DPQA can compile 90-qubit benchmarks in one day, whereas Enola compiles 100 qubit circuits with higher fidelity in a minute.

Fig.~\ref{fig:runtime} exhibits the runtime of Enola with sortIS and windowIS (default) on larger benchmarks, up to 10,000 qubits.
Note that this is a log-log plot.
Different colors inside each bar provides the portion of time spent on different tasks.
The scheduling is extremely fast, invisible in the plot.
For benchmarks smaller than 1,000 qubits, the runtime is dominated by the placement.
Although the placement scales in $O(n)$, the constant factor is larger than routing.
Later on, the routing portion becomes dominant due to a higher asymptotic: sortIS takes $O(n^4)$ time and windowIS takes $O(n^2\log n)$ time with a constant window size.
At 10,000 qubits, the sortIS approach took 1.22e4 seconds, i.e., about 3.4 hours; the windowIS approach took 1.50e3 seconds, i.e., about 25 minutes.
From the data, the runtime scaling of windowIS roughly follows $O(n^2 
\log n)$: increasing the number qubits by 10x from 1,000 to 10,000, the runtime increases by 55x from 18.8 seconds to 1.04e3 seconds.

\subsection{Potential Impact of Multiple AODs}

As noted in Ref.~\cite{wang2024atomique}, it is feasible to equip DPQA with multiple AODs to facilitate the simultaneous movement of different sets of atoms.
The AOD ordering constraints apply only to atoms within the same AOD, allowing multiple AODs to further enhance parallelism in the routing task.
Since this multi-AOD configuration has not been experimentally demonstrated, we provide only a rough estimation of its potential impact using a simple round-robin assignment strategy.
For example, if the routing stage identifies four compatible sets of moves, $M_0$ to $M_3$, and there are two AODs available, we can utilize AOD\_0 to execute $M_0$ and AOD\_1 to execute $M_1$, simultaneously.
The duration of this `trunk' of movements ($M_0$ and $M_1$), is determined by the longer of the two durations.
Subsequently, as the next trunk, AOD\_0 implements $M_2$ while AOD\_1 executes $M_3$.
The total duration of routing is then the cumulative duration of all trunks.
This parallelism among multiple AODs reduces the overall computation time, leading to lower decoherence error and improved overall fidelity.
On the 90-qubit QAOA benchmarks, we observe that utilizing two, three, and four AODs enhances the overall fidelity by factors of 1.54x, 1.79x, and 1.94x, respectively.

\section{Related Work}\label{sec:related}
We have introduced the three existing works that can be directly compared with Enola: OLSQ-DPQA~\cite{iccad22-olsq-raa, tan2023compiling}, Q-Pilot~\cite{wang2023qpilot}, and Atomique~\cite{wang2024atomique}.
In this section, we discuss other related works.

A key hardware assumption in our work is that the Rydberg laser globally excites all qubits.
Although individually addressed Rydberg lasers have been demonstrated~\cite{nature22-graham-atom-array}, the scale and operation fidelity of such platforms are significantly lower than that of the global approach~\cite{nature22-lukim-bluvstein-atom-array}.
With individual addressability, qubits that are idling during a Rydberg exposure will not experience the same noise level as those involved in two-qubit gates.
Furthermore, moving atoms may be unnecessary when individual addressability is available, as qubits do not need to avoid each other if they are not participating in two-qubit gates during a Rydberg exposure.
In this scenario, the optimization objective would shift from minimizing the number of Rydberg stages to minimizing the number of gates.

Consequently, the qubits can be routed logically using SWAP gates, similar to the methods employed in quantum computing platforms with a fixed coupling graph, such as superconducting circuits.
Due to the fundamental differences in optimization objectives, most of the mentioned previous approaches~\cite{circuit-placement, fan_QLSML_2022, asplos19-li-ding-xie-sabre-mapping, dac19-wille-burgholzer-zulehner-mapping-minimal-swaph, iccad21-tan-cong-qubit-mapping-absorption, huang2022reinforcement, li2023single_qubit_gate_matter_for_qls, park2022fsqm, asplos21-zhang-hayes-qiu-jin-chen-zhang-time-optimal-mapping, micro22-tannu-maxsat-qubit-mapping, tc20-tan-cong-optimality-layout-queko, iccad20-tan-cong-optimal-layout-synthesis, dac23-olsq2, wu_robust_2022,  ShaikvdP2023_qls_as_classical_planning} does not directly apply to the DPQA compilation problem considered in this work.

Baker et al.~\cite{isca21-baker-litteken-duckering-hoffmann-bernien-chong-long-distance-neutral} investigated layout synthesis for a fixed atom array with individually addressed Rydberg lasers.
Li et al.~\cite{ict-neutral-atom} further considered the detailed durations for different gates in the scheduling process under this hardware setting.
Patel et al.~\cite{isca22-patel-silver-tiwari-geyser-neutral} proposed a method for logic resynthesis to leverage three-qubit gates available on neutral atoms, with the SWAPs for routing qubits sometimes becoming a `free lunch' after the resynthesis.

Several works utilized movement capabilities even when assuming an individually addressed Rydberg laser.
Brandhofer et al.~\cite{iccad21-brandhoher-buchler-polian-optimal-mapping-atoms} targeted an architecture with a more restricted type of movement known as `1D displacement.'
Schmid et al.~\cite{schmid2023hybrid} proposed a combination of SWAP and AOD movements for routing qubits.
Again, due to the differing hardware assumptions, these previous approaches cannot be directly compared with the work presented here.

\section{Conclusion and Future Direction}\label{sec:conclusion}
In this paper, we formulated three tasks in the layout synthesis for DPQA: scheduling, placement, and routing.
We presented efficient solutions to all of them and integrated these solutions in our compiler, Enola.
Most notably, because the scheduling is based on the Misra-Gries edge coloring algorithm, Enola generates provably near-optimal number of two-qubit gate stages.
Our placement based on simulated annealing and routing based on independent set also proves effective.
This paper leads to a few promising directions.

1) Synergy between the three tasks.
In this work, our compilation is conducted in a single pass that encompasses scheduling, placement, and routing.
However, it may be advantageous to implement further multi-pass optimizations or iterative refinement of solutions for these problems.
For example, currently the edge coloring assigns two-qubit gates to stages but the order of these stages is random.
Exploring this ordering resembles placement-driven scheduling~\cite{guo2021autobridge} and may further decrease the total movement time.
Further improvement is also possible via routability-based placement~\cite{li2007routability} because the current placement only reduces the total Euclidean distance of two-qubit gates without considering whether the moves corresponding to these gates are compatible.

2) More detailed formulation of the routing problem.
For example, the current notion of compatibility considers the entirety of movements.
However, a move may not entirely be compatible with another move but becomes compatible after it progresses past a certain portion.
Exploring these opportunities may also require innovation in the lower-level instruction set.
Separately, the NP-hardness of routing justifies further exploration in solver-based methods.
Although these methods cannot solve large-scale problems, it is still valuable to pursue optimal solutions for critical and frequent subroutines or a coarsened solution in a multilevel flow.

3) Application-specific compilation.
This work achieves the highest improvement over previous state of the art on commutation groups of two-qubit gates.
However, there are commutation relations on higher level structures like Pauli string unitaries in quantum simulation applications~\cite{paulihedral} that lies out of the scope of this paper.
Another example is the syndrome extraction for quantum error correcting codes, especially good quantum low-density parity-check codes~\cite{qldpc, xu2024constant}.
Although these circuits are not commutation groups, there is flexibility in the ordering of two-qubit gates.

4) Adaptation to hardware capabilities and co-design.
For example, it is possible to trap atoms in multiple 2D planes, making the architecture 3D \cite{atom3d}.
Another example is DPQA with separate storage and entanglement zones~\cite{bluvstein2023logical} where the Rydberg laser only illuminates the entanglement zone.
This work is still useful to handle what happens inside the entanglement zone, but a higher-level framework of shuttling qubits between the zones should be developed.
An efficient and effective compiler such as Enola also aids in the hardware co-design by providing improvement estimations.

\section*{Acknowledgment}
This work is funded by NSF grant 2313083.
The authors would like to thank Dolev Bluvstein, Harry Hengyun Zhou and Prof. Mikhail D. Lukin for valuable discussions.

\bibliographystyle{ACM-Reference-Format}
\bibliography{newrefs}

\end{document}